# Fabry-Pérot Huygens' Metasurfaces:
# On Homogenization of Electrically Thick Composites

Sherman W. Marcus and Ariel Epstein

Andrew and Erna Viterbi Faculty of Electrical Engineering, Technion - Israel Institute of Technology, Haifa 3200003, Israel

Realization of the anomalous refraction effects predicted by Huygens' metasurfaces (HMS) have required tedious and time-consuming trial-and-error numerical full-wave computations. It is shown herein that these requirements can be alleviated for transverse magnetic (TM) propagation by a periodic dielectric-based HMS consisting of an electrically thick array of cascaded Fabry-Pérot etalons. This "Fabry-Pérot HMS" (FP-HMS) is easily designed to mimic the local scattering coefficients of a standard zero-thickness HMS (ZT-HMS) which, according to homogenization theory, should result in the desired anomalous refraction. To probe the characteristics of this practical FP-HMS, a method based on Floquet-Bloch (FB) analysis is derived for predicting the fields scattered from it for arbitrary angles of incidence. This method produces simple closed-form solutions for the FB wave amplitudes and the resulting fields are shown to agree well with full-wave simulations. These predictions and full-wave simulations verify the applicability of homogenization and scattering properties of zero-thickness HMS's to thick structures. They also verify the proposed semi-analytical microscopic design procedure for such structures, offering an effective alternative path to implementation of theoretically envisioned intricate field manipulating devices.

## I. INTRODUCTION

Anomalous refraction effects have been predicted for scattering of electromagnetic waves across Huygens' metasurfaces (HMS) [1]-[3]. These are virtual zero-thickness surfaces that are characterized by their surface electric impedance $Z_{se}$ and surface magnetic admittance $Y_{sm}$, while the relationship between these surface properties and the fields along the surface are encompassed in generalized sheet transition conditions (GSTCs) [4]-[11]. The "microscopic design" of actual devices with these wave manipulation properties has necessitated characterization of subwavelength elements ("meta-atoms") exhibiting the required wave responses, with these responses determined by full-wave scattering simulations [10]. The responses as a function of the element characteristics (size, shape, material), and their relationship to the metasurface properties $Z_{se}$ and $Y_{sm}$, would then be tabulated in a look-up table which would often require refinement through optimization [12]-[20]. Finally, entries in this table would be chosen which provided the closest value of the required $Z_{se}$ and $Y_{sm}$ for each meta-atom along the HMS; this could often result in a tedious and less-than-precise design procedure.

In contrast to the virtual zero-thickness surfaces, and in contrast to the actual devices that require tedious design procedures, we propose in this paper a simple, realistic, straightforward, semi-analytical "microscopic" design for the HMS, which eliminates the need for full-wave computations and look-up tables. It employs dielectric slabs within parallel-plate waveguides which serve as meta-atoms, making it conceptually simple both in its geometry and in the material of which it is composed, and suitable for fabrication with the aid, say, of additive manufacturing (3D printing) techniques. The electrically thick structure will be referred to as a Fabry-Pérot HMS (FP-HMS) since it consists of an array of cascaded etalons with two reflecting boundaries which are tuned to provide the desired transmission amplitude and phase [21].

Not only is the design of the proposed FP-HMS conceptually simple, but it is this simplicity that permits its macroscopic characteristics to be expressed in closed-form. These closed-form expressions not only include those for propagated fields due to the incident wave for which the structure was designed, but include expressions for *all* fields, both propagated and evanescent, both within and outside the structure, and for *all* angles of incidence. The ability to obtain such solutions in closed form for practical physical geometries is somewhat unique.

It should be noted that other studies dealing with scattering from planar formations for various functionalities at optical frequencies have used geometries with some similarities to the one considered here, but have had to resort to full-wave solutions for field calculations. These include high contrast diffraction gratings [22], dielectric metasurfaces based on geometric phase [23]-[26], and resonant meta-atoms [27]-[30], to name a few. In contrast to these other studies which did not aim at a Huygens response, the goal in this paper is to provide a general inhomogeneous configuration, and an accompanying analytical framework, that is capable of beam deflection and radiation molding not accessible with simple periodic structures. In particular, the meta-atom synthesis generally requires optimization in full-wave solvers; this is avoided herein by harnessing conducting walls to separate adjacent unit cells, which is more realistic in the microwave regime, enabling the rigorous microscopic design. Such a separation has previously been utilized in producing "extraordinary transmission" phenomena [31]-[33], and for controlling reflected modes, from the early microwave reflectarray antennas [34] to concurrent anomalous reflection acoustic metasurfaces [35], wherein the phase variation is attained by varying the parallel-plate waveguide length. For transmit-mode applications, however, a different approach, such as the one introduced here, is required, as mere variation of the waveguide length cannot effectively modulate the phase of the transmission coefficient.

The "canonical" HMS problem for anomalous refraction that will be studied herein involves a plane wave that is incident on the upper face of the HMS at one angle, and is transmitted across the HMS as a plane wave at another angle with little specular reflection. Section II describes the basis and design for a simple, straightforward, realistic HMS based on Fabry-Pérot principles, while Section III provides closed-form expressions for the fields from this FP-HMS which are used for analyzing the scattering characteristics of the FP-HMS. In particular, the well-known functional forms of $Z_{se}$ and $Y_{sm}$ of an *abstract* zero-thickness HMS (ZT-HMS) that produces this anomalous refraction effect are reviewed in Section II, along with relations between these $Z_{se}$ and $Y_{sm}$ functions and local transmission and reflection coefficients along the ZT-HMS [3],[36]. It is these coefficients that will be mimicked in an actual design of the FP-HMS that will also be described in Section II and which, according to homogenization theory, should reproduce the scattering predicted by the ZT-HMS.

Section III, which is dedicated to analysis of the fields scattered from the FP-HMS, employs Floquet-Bloch formalism to formulate these fields in terms of an infinite system of linear equations which is then solved in closed form. Results are provided in Section IV, where full agreement is demonstrated for scattering from the proposed FP-HMS predicted by the closed-form solution, full-wave solutions, and, for incident angles of interest, solutions reported previously for the ZT-HMS [37]. Besides validating homogenization for electrically thick devices such as the proposed FP-HMS, the presented systematic synthesis and analysis schemes are expected to open an efficient and reliable alternative path for practical realization of advanced beam-manipulating HMSs.

## II. SYNTHESIS

In this section, the expressions for the ZT_HMS characteristics $Z_{se}$ and $Y_{sm}$ are reviewed which produce anomalous refraction, and the corresponding (continuous) distribution of the local transmission and reflections coefficients $T$ and $R$ along its surface are found. A simple, straightforward method based on Fabry-Perot principles is then described for designing a discretely-varying structure which approximates these $T$ and $R$ distributions.

### 1. ZT-HMS and FP-HMS Transmission Coefficients

Consider a ZT-HMS that coincides with the *x-z* plane and is illuminated by a transverse magnetic (TM) polarized plane wave ($H_x=H_y=E_z=0$) at an incidence angle $\theta_{inc}$ relative to the normal to the surface, as shown in Fig. 1.

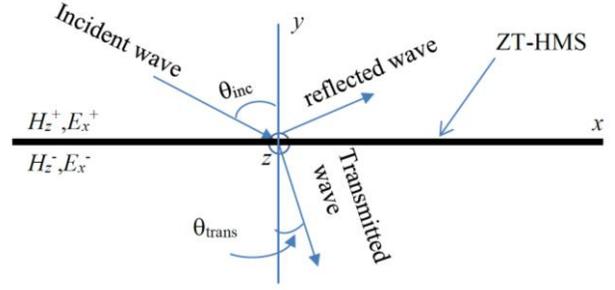

FIG 1. A zero-thickness Huygens' metasurface illuminated by a plane wave. The surface with periodic properties produces a discontinuity in the field components that are tangent to the surface. The *H*-field values above and below the surface are $H_z^+$ and $H_z^-$, respectively, and the *x*-components of the *E*-field above and below the surface are $E_x^+$ and $E_x^-$, respectively. For the $Z_{se}$ and $Y_{sm}$ given in Eqs. (3) and (4), and for an incidence angle equal to the "designated" incidence angle (see below), only a specularly reflected wave and a single transmitted wave will be produced as shown.

The ZT-HMS is characterized by the periodic electric surface impedance $Z_{se}(x)$ and the periodic magnetic surface admittance $Y_{sm}(x)$ which define the GSTCs [1], [4], [10],

$$Z_{se}(x)[H_z^-(x) - H_z^+(x)] = \frac{1}{2}[E_x^-(x) + E_x^+(x)], \quad (1)$$

$$Y_{sm}(x)[E_x^-(x) - E_x^+(x)] = \frac{1}{2}[H_z^-(x) + H_z^+(x)], \quad (2)$$

where the fields $H_z^+$, $H_z^-$, $E_x^+$, $E_x^-$ are the field components tangential to the upper surface (+ superscript) and the lower surface (− superscript).

It is desired to design the ZT-HMS to refract a wave from an incident direction $\theta_{inc}$ to a transmission direction $\theta_{trans}$. This functionality was addressed rigorously in [3] and [36], harnessing the HMS design rules to analytically resolve the passive lossless surface constituents required to implement it. For a period $d$, these were found to require

$$Z_{se}(x) = \frac{i}{2} Z \cos\theta_{trans} \cot\frac{kx\Delta_{\sin}}{2}, \quad 0 \leq x \leq d, \quad (3)$$

$$Y_{sm}(x) = \frac{i}{2} \frac{1}{Z\cos\theta_{trans}} \cot\frac{kx\Delta_{\sin}}{2}, \quad 0 \leq x \leq d, \quad (4)$$

where

$$\Delta_{\sin} = \sin\theta_{trans} - \sin\theta_{inc}, \quad d = \frac{\lambda}{|\Delta_{\sin}|}, \quad (5)$$

an $e^{-i\omega t}$ time dependence is implied, $k=\omega(\varepsilon_0\mu_0)^{1/2}=2\pi/\lambda$ is the wave number in free space, $\lambda$ is the wavelength, $\varepsilon_0$, $\mu_0$ are the permittivity and permeability of free space, $\omega=2\pi f$ is the angular velocity of the wave, $f$ is the frequency, and $Z=(\mu_0/\varepsilon_0)^{1/2}$ is the free space impedance. It is desired to design the FP-HMS to emulate this same anomalous refraction functionality. To facilitate such a passive lossless design, some (generally minor) specular reflection is required in addition to the desired transmitted mode, stemming from the need for local impedance equalization [36].

The procedure for approximating the abstract boundary conditions in Eqs. (3) and (4) by a physical structure (i.e. the FP-HMS) involves expressing the $Z_{se}(x)$ and $Y_{sm}(x)$ functions in terms of local transmission and reflection coefficients $T(x)$ and $R(x)$ along the ZT-HMS for a normally incident plane wave [38]. This is accomplished by discretizing the sheet along $x$, and for each particular value $x=x_p$ of $x$, considering a surface characterized entirely by $Z_{se}(x_p)$, $Y_{sm}(x_p)$ (i.e. assuming local homogeneity about $x=x_p$) [10]. Although this procedure can be applied for $Z_{se}(x)$ and $Y_{sm}(x)$ obtained using any design angles $\theta_{inc}$ and $\theta_{trans}$, a particularly simple result for $T(x)$ and $R(x)$ is obtained when $\theta_{trans}=0$ in (3) to (5):

$$R_p \equiv R(x_p) = 0,\; T_p \equiv T(x_p) = e^{i\phi_p} = e^{-ikx_p \Delta_{\sin}} = e^{-i2\pi x_p/d}. \quad (6)$$

In this manuscript, we will therefore restrict ourselves to cases for which $\theta_{trans}=0$. (We envision future study of other values for $\theta_{trans}$ for which the expressions for $T$ and $R$ would be more general.) Thus, from Eq. (6), when the FP-HMS is illuminated by a normally incident wave, each of its discretized intervals should have a vanishing reflection coefficient, and a transmission coefficient with unit magnitude and constant phase gradient. Since the parameter $x_p$ can be any value of $x$ between 0 and $d$, the phase of the transmission coefficient will take on a corresponding value between 0 and $-2\pi$. It is these reflection and transmission coefficients that will be emulated in the practical FP-HMS considered below.

## 2. FP-HMS Design

In order to emulate the ZT-HMS of Eqs. (3) and (4), the periodic transmission and reflection coefficients in Eq. (6) will now be implemented in a finite thickness slab-shaped body located between $y=0$ and $y=-h$, with discrete divisions of small width $\Delta \ll \lambda$ along the $x$-direction as shown in Fig. 2. If there are $N_{wg}$ such divisions within a period $d$, then the desired transmission coefficient of the $p$th division would be given by

$$T_p = T(x_p) = e^{-i2\pi x_p/d},\; x_p = \left(p - \frac{1}{2}\right)\Delta,\; p = 1, 2, ..., N_{wg}. \quad (7)$$

$T_p$ should characterize the wave propagation in the $p$th division without being affected by the propagation in adjacent divisions. This can be attained by isolating each division, which is accomplished by enclosing it within perfectly (electrically) conducting plates parallel to the $y$-$z$ plane as shown in Fig. 2. The entire FP-HMS therefore consists of an array of parallel plate waveguides. Each "waveguide" $p$ must now be filled in a manner which would produce transmission coefficients $T_p$ which satisfy Eq. (7) for $y$-directed propagation from one end of the waveguide to the other.

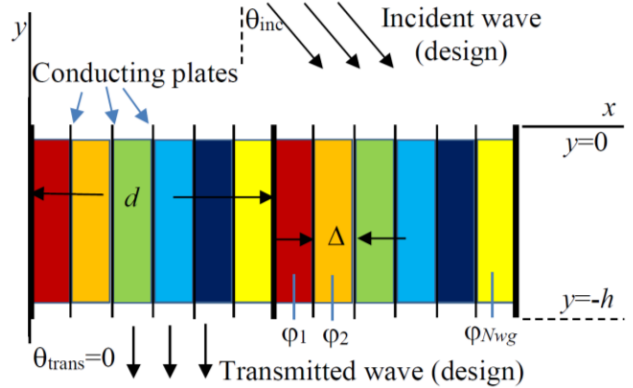

FIG. 2. A slab formed by a periodic array (period $d$) of narrow waveguides of width $\Delta$. The slab boundaries are at $y=0$ and $y=-h$. Each waveguide $p$ contains material which provides it with a transmission coefficient $T_p$ for propagation in the $y$-direction.

In conformance with the method used to derive Eq. (6), the material distribution which will result in the proper $T_p$ within waveguide $p$ is determined below for waves that propagate normal to layers of homogeneous slabs (Fig. 3). For the assumed TM polarization, the $E$-field is normal to the conducting plates so that these plates have no effect on the propagation and can be ignored. (This is, in effect, the rationale for employing TM polarization.) The possibility of applying impenetrable impedance surfaces (e.g., as in [39]) to support both TM and TE polarizations will be explored in the future. Finally, since the waveguides are narrow, the propagation direction can be assumed to be normal to the layers even when the incident wave impinges obliquely on the waveguide array since within the waveguide only the zeroth order waveguide mode – which propagates in the $y$-direction – will contribute to the field [40].

### a. $T$ and $R$ for a Three-Layer Medium

The fields in a medium consisting of $N$ layers will now be found for a plane wave normally incident on it from above, as shown in Fig. 3. Each layer is characterized by its wave number $k_i=k_0(\varepsilon_{ri})^{1/2}$ and its lower boundary $y=y_i$, where the 0 subscript refers to the incidence region, $k_0$ is the free space wave number, $\varepsilon_{ri}$ is the relative permittivity of layer $i$, and the relative permeability of each layer is unity. For this case, both the $\mathbf{H}$ and the $\mathbf{E}$ fields will be tangent to the boundaries, so that in layer $i$, $\mathbf{H}_i=H_i\hat{\mathbf{z}}$, $\mathbf{E}_i=E_i\hat{\mathbf{x}}$. The fields in each region satisfy the Helmholtz equation

$$\frac{\partial^2 H_i}{\partial y^2} + k_i^2 H_i = 0, \quad y_{i-1} \leq y \leq y_i. \quad (8)$$

the general solution of which is [41]

$$H_i(y) = A_i e^{-ik_i y} + B_i e^{ik_i y}, \quad 0 \leq i \leq N+1 \quad (9)$$

so that

$$E_i(y) = Z_i\left(A_i e^{-ik_i y} - B_i e^{ik_i y}\right), \quad 0 \leq i \leq N+1 \quad (10)$$

where $i=N+1$ indicates the bottom (transmission) layer, and $Z_i=Z/\varepsilon_{ri}^{1/2}$ is the wave impedance in layer $i$. In the incidence region, $A_0$ is assumed known since it characterizes the incident wave. In the transmission region, $B_{N+1}=0$ to satisfy the radiation condition. All the other $A_i$, $B_i$ are $2(N+1)$ unknowns which can be readily determined in terms of $A_0$ by solving the linear system of equations formed from the $2N+2$ boundary conditions $H_{i-1}(y_{i-1})=H_i(y_{i-1})$, $E_{i-1}(y_{i-1})=E_i(y_{i-1})$, $1\leq i \leq N+1$.

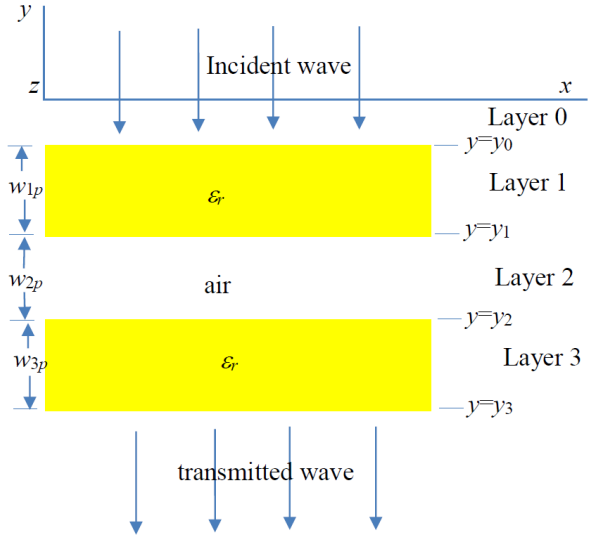

FIG. 3. Layered configuration for determining the amplitudes of the transmitted and reflected waves.

For the case shown in Fig. 3, $N=3$, $y_1=y_0-w_{1p}$, $y_2=y_0-(w_{1p}+w_{2p})$, $y_3=y_0-(w_{1p}+w_{2p}+w_{3p})$, $\varepsilon_{r1}=\varepsilon_{r3}=\varepsilon_r$, $\varepsilon_{r0}=\varepsilon_{r2}=\varepsilon_{r4}=1$, $Z_0=Z_2=Z_4=Z$, $Z_1=Z_3=Z/\varepsilon_r^{1/2}$, $k_0=k_2=k_4=k$, $k_1=k_3=k\varepsilon_r^{1/2}$. The reflection and transmission coefficients are defined as

$$R_p=B_0/A_0, \quad T_p=A_{N+1}/A_0=A_4/A_0, \quad (11)$$

where $B_0$ and $A_4$ can be easily obtained as solutions of the above linear system. We are generally limited to specific materials, so that $\varepsilon_r$ can be assumed given and the only variables are $w_{1p}$, $w_{2p}$ and $w_{3p}$. Thus, for any combination of these variables, the value of the complex transmission coefficient $T_p$ can be found. But it is known that HMS meta-atoms possess two degrees of freedom, and that they can be represented by a symmetric arrangement [2],[42]. To conform with this geometry, we will impose $w_{1p}=w_{3p}$, so that

$$T_p = T_p(w_{1p}, w_{2p}), \quad (12)$$

where $0\leq|T_p|\leq 1$ for passive media, and the two degrees of freedom are manifested by the two unknowns $w_{1p}$ and $w_{2p}$.

**b. Solutions for Layer Thicknesses**

We solve the nonlinear equation in Eq. (12) using the Matlab function lsqnonlin to retrieve the values of $w_{1p}$ and $w_{2p}$ for the desired $T_p$ of Eq. (7) for all values of $x_p$ [43]. Dielectric material characterized by $\varepsilon r=16$ (index of refraction = 4) was chosen for use in all the waveguides of our design, and thus for use in Eq. (12). Results for the desired thicknesses $w_1$ and $w_2$ are shown in Fig. 4 as a function of $\phi_p$. The $w_1$ and $w_2$ functions of course are periodic, but not continuous. Since any $2\pi$ region of $\phi_p$ could have been chosen to display $w_1$ and $w_2$, the discontinuity at $\phi_p=60°$ could have been placed at either end of the period. It was purposely left unaltered to emphasize its existence.

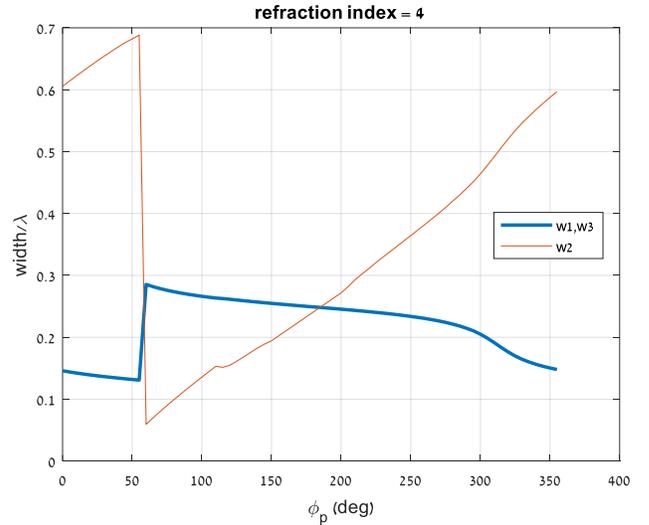

FIG. 4. The widths $w_1$ of layer 1 and layer 3 (thick curve), and $w_2$ of layer 2 (thin curve), as a function of phase angle $\phi_p$ for $|T_{targ}|=1$ as obtained from the solution of Eq. (12).

When solving Eq. (12), singular behavior of the air-gap function $w_2$ was observed in the region $\phi_p=180°$. Fig. 4 reflects the $w_2$ values after having been smoothed in that region. This smoothing was found to have only a negligible effect on the transmission coefficient.

Fig. 5 illustrates three periods of the entire array for the case of $\theta_{inc}=80°$, $\theta_{trans}=0$ when the material is placed symmetrically within the waveguides about $y=-h/2$. All the conducting plates are the same height $h$ to maintain the semblance of a slab with boundaries at $y=0$ and $y=-h$, so that

$h \geq \max_p \left(2w_{1p} + w_{2p}\right)$. We stress that although the unit cells feature five layers, the two outermost air regions do not contribute to the local transmission coefficient; hence, the three-layer model discussed and analyzed in Section II.2.a is valid here as well. Just as in the ZT-HMS case, the period $d$ of the FP-HMS shown in Fig. 5 is determined by the designated incidence and transmission angles $\theta_{inc}$ and $\theta_{trans}$ in accordance with Eq. (5).

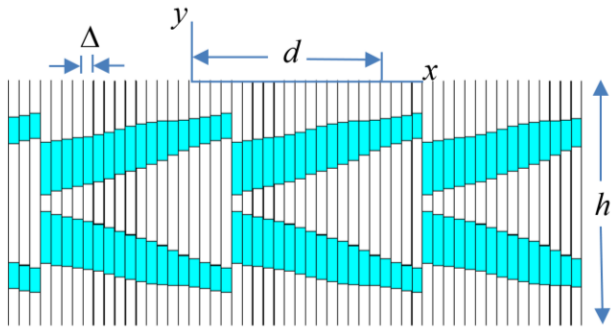

FIG. 5. The material widths shown in Fig. 4 arranged with vertical symmetry, and distributed discretely over three period-lengths for $d=1.015\lambda$ (i.e. $\theta_{inc}=80°$, $\theta_{trans}=0$), $h=1.3\lambda$, $\Delta=d/18$. The material portion of each waveguide is colored, the air portion is white.

## III. ANALYSIS

In the previous section, Eq. (7) was used as the basis for detailed design of the FP-HMS. In this section, it will be used as the basis for an analytical model composed of the parallel-plate waveguides in Fig. 2 within which the field is characterized only by $T(x)$. This model will allow us to explore the response of the FP-HMS when excited by a plane wave at angles of incidence $\psi_{inc}$ that differ from the incidence angle $\theta_{inc}$ for which the FP-HMS was designed [37]. For such cases, since the "grating momentum" is fixed, the angle $\psi_{trans}$ of the resulting transmitted wave would generally differ from the design value $\theta_{trans}$. Comparison of these predictions with those in [37] for the ZT-HMS will allow us to evaluate the performance of the FP-HMS as a homogenized metasurface, an interesting concept considering the large electrical thickness of the physical structure.

### 1. Formulation

Assuming TM propagation and no field or geometry variation in the $z$ direction, the magnetic field can be written $\mathbf{H}(x,y) = H(x,y)\hat{\mathbf{z}}$, where $H(x,y)$ satisfies the Helmholtz equation

$$\frac{\partial^2 H(x,y)}{\partial x^2} + \frac{\partial^2 H(x,y)}{\partial y^2} + k^2 H(x,y) = 0. \quad (13)$$

Referring to Fig. 2, the fields will have different expressions above, below and within the FP-HMS:

$$H(x,y) = \begin{cases} H_{inc}(x,y) + H_{ref}(x,y), & y > 0, \\ H_2(x,y), & -h \leq y \leq 0, \\ H_{trans}(x,y), & y < -h, \end{cases} \quad (14)$$

where the incident wave of unit amplitude is given by

$$\mathbf{H}_{inc}(x,y) = \hat{\mathbf{z}} e^{ikx\sin\psi_{inc} - iky\cos\psi_{inc}}. \quad (15)$$

The reflected field $H_{ref}$ and the transmitted field $H_{trans}$ are defined in accordance with Floquet-Bloch (FB) theory as super-positions of reflected plane waves and transmitted plane waves, respectively:

$$H_{ref} = \sum_{n=-\infty}^{\infty} \rho_n e^{i\alpha_n x + i\beta_n y}, \quad H_{trans} = \sum_{n=-\infty}^{\infty} \tau_n e^{i\alpha_n x - i\beta_n(y+h)}, \quad (16)$$

where

$$\alpha_n = k\sin\psi_{inc} + \frac{2n\pi}{d}, \quad \beta_n = \sqrt{k^2 - \alpha_n^2}, \quad (17)$$

$\rho_n$ and $\tau_n$ are the initially unknown amplitudes of the reflected and transmitted waves, respectively, and the branch of $\beta_n$ is chosen to satisfy the radiation condition for $|y|\to\infty$.

To allow homogenization, we require that the width $\Delta$ of each parallel plate waveguide be small, $\Delta \ll \lambda$ [4]. As a result, only the zeroth order waveguide mode will propagate within it. The field $H_2$ within the FP-HMS must be written separately for each of the $N_{wg}$ waveguides within the period $d$. If waveguide $p$ were empty, the field within it could be written as the sum of an upward wave and a downward wave: $H_{2p} = \sigma_{pa} e^{-iky} + \sigma_{pb} e^{iky}$, where $\sigma_{pa}$ and $\sigma_{pb}$ are the respective wave amplitudes. Although it is *not* empty, the transmission coefficient $T_p$ has already been found in Eq. (12). Although $T_p$ characterizes the field only at the opposite end of the waveguide relative to the incident wave, this field is sufficient for formulating boundary conditions which only require the field values at the waveguide extrema. The field within waveguide $p$ may therefore be written using separate expressions near its top and near its bottom:

$$H_2(x,y) = H_{2p}(y) = \begin{cases} \sigma_{pa} e^{-iky} + T_p \sigma_{pb} e^{iky}, & y \to 0^-, \\ T_p \sigma_{pa} e^{-iky} + \sigma_{pb} e^{iky}, & y \to -h^+, \end{cases} \quad (18)$$

where $1 \leq p \leq N_{wg}$, and the amplitudes $\sigma_{pa}$ and $\sigma_{pb}$ are to be determined. $H_{2p}(y)$ depends on $x$ through $p$ which is a discrete function of $x$: $p = 1 + \text{int}(x/\Delta)$. To understand the right side of Eq. (18), consider the field near the top of the waveguide ($y \to 0^-$). The "incident" downward wave there, $\sigma_{pa} e^{-iky}$, will have the same form as that for an empty waveguide. The upward wave there started as an "incident" wave $\sigma_{pb} e^{iky}$ at the bottom, and reached the top after propagating through the dielectric loading for which the transmission coefficient is $T_p$. The amplitude of the upward wave at the top is therefore $T_p\sigma_{pb}e^{iky}$, as given in the $y\to 0^-$ expression in Eq. (18). The $y\to -h^+$ expression can be understood in the same manner.

It should be emphasized that the fact that Eq. (18) is based on perfect transmission within each waveguide does *not* imply perfect transmission through the entire FP-HMS, even for a normally incident plane wave. This is because of an impedance mismatch introduced as a result of the lateral inhomogeneity of the HMS design [10], [11], [36], [37], [38], [42], [44].

The field definitions in Eqs. (14) to (18) are used in Appendix A to determine the boundary conditions which must be satisfied. These lead to (A9) and (A10):

$$-\sum_{n=-\infty}^{\infty}\rho_n C_n e^{i\left(\frac{2n\pi}{d}\right)x}+e^{-i2\pi x/d}e^{ikh}\sum_{n=-\infty}^{\infty}\tau_n S_n e^{i\left(\frac{2n\pi}{d}\right)x}=S_0, \quad (19)$$

$$-\sum_{n=-\infty}^{\infty}\rho_n S_n e^{i\left(\frac{2n\pi}{d}\right)x}+e^{i2\pi x/d}e^{-ikh}\sum_{n=-\infty}^{\infty}\tau_n C_n e^{i\left(\frac{2n\pi}{d}\right)x}=C_0, \quad (20)$$

where

$$S_n=\frac{1}{2}(1-\gamma_n), \; C_n=\frac{1}{2}(1+\gamma_n), \; \gamma_n=\frac{\beta_n}{k}. \quad (21)$$

It is worth noting that $\gamma_n$ is the wave impedance ratio between the $n$th FB mode and a normally incident plane wave, while the ratio $\Gamma_n=S_n/C_n$ corresponds to the associated Fresnel reflection coefficient [37]. From Eqs. (21) and (17), the following identities are useful:

$$C_n+S_n=1, \; C_n^2-S_n^2=C_n-S_n=\gamma_n, \quad (22)$$

$$\gamma_0=\cos\psi_{inc}, \; C_0=\cos^2(\psi_{inc}/2), \; S_0=\sin^2(\psi_{inc}/2), \quad (23)$$

$$\gamma_{-1}=\cos\psi_{trans}, \; C_{-1}=\cos^2(\psi_{trans}/2), \; S_{-1}=\sin^2(\psi_{trans}/2). \quad (24)$$

For later use, when $\psi_{inc}=\theta_{inc}$, then $\psi_{trans}=\theta_{trans}$, and $\gamma_{-1}=\cos\theta_{trans}$. When $\psi_{trans}=\theta_{trans}=0$, then

$$\gamma_{-1}=C_{-1}=1, \; S_{-1}=0. \quad (25)$$

Once Eqs. (19) and (20) are solved for the $\rho_n$ and $\tau_n$, these may be used in Eqs. (A5) and (A6) to yield the wave coefficients within the FP-HMS:

$$\sigma_{pa}=\sigma_a(x)=C_0 e^{i\alpha_0 x}+\sum_{n=-\infty}^{\infty}S_n\rho_n e^{i\alpha_n x}, \quad (26)$$

$$\sigma_{pb}=\sigma_b(x)=e^{ikh}\sum_{n=-\infty}^{\infty}S_n\tau_n e^{i\alpha_n x}. \quad (27)$$

Although Eq. (18) sufficed for describing the fields within the FP-HMS for purposes of formulating boundary conditions, it does not prescribe the fields within each waveguide $p$. However, consistent with the uniqueness theorem [45], such fields *can* be found by realizing from the form of Eq. (18) that $\sigma_{pa}e^{-iky}$ plays the role of a downward incident wave from above the FP-HMS, and $\sigma_{pb}e^{iky}$ plays the role of an upward incident wave from below. These waves propagate through the dielectric layers which produced the transmission coefficient $T_p$ appearing in Eq. (18). The $\sigma_{pa}$ found in Eq. (26) therefore represents the amplitude $A_0$ of the wave incident on the three-layer medium of Fig. 3 from above. The field due to this wave in medium $i$ is given in Eq. (9) as

$$H_{2pai}(y)=A_i(A_0=\sigma_{pa})e^{-ik_i y}+B_i(A_0=\sigma_{pa})e^{ik_i y}, \quad (28)$$

with the coefficients $A_i$, $B_i$ obtained by solving the linear system represented by Eqs. (9) and (10) and discussed in Section II.2. This provides the contribution of $\sigma_{pa}$ to the fields in each layer $i$ throughout waveguide $p$. The contribution to these fields of the upward wave characterized by $\sigma_{pb}$ is found in the same manner: $\sigma_{pb}$, as determined in Eq. (27), also represents the incident amplitude $A_0$, and since the geometry is symmetric about $y=-h/2$, the solution for the fields will be the same except for a factor related to the ratio of $\sigma_{pb}$ to $\sigma_{pa}$:

$$H_{2pbi}(y)=A_i(A_0=\sigma_{pb})e^{-ik_i y}+B_i(A_0=\sigma_{pb})e^{ik_i y}, \quad (29)$$

When adding the "$\sigma_{pb}$-fields" to the "$\sigma_{pa}$-fields", care should be taken to account for the difference in direction of the propagation:

$$H_{2pi}(y)=H_{2pai}(y)+H_{2pb,N+1-i}(-h+y). \quad (30)$$

Eq. (30), which prescribes the fields inside each waveguide, supplements the solution of Eqs. (19) and (20) which prescribes the fields above and below the waveguides, thereby providing a complete formulation of the scattering problem everywhere in space.

### 2. Solution

Eqs. (19) and (20) represent two equations in the two sets of unknowns $\rho_n$, $\tau_n$. They are generally solved by projecting each term in both equations into the Rayleigh basis $e^{2im\pi x/d}$ [46]. This would provide two equations for each integer index $m$. The number of such $m$-values employed would be such as to provide the same number of equations as the number of unknowns.

Equating coefficients of $e^{2\pi inx/d}$ for the same value of $n$ in Eqs. (19) and (20) leads to

$$-C_m\rho_m+e^{ikh}S_{m+1}\tau_{m+1}=S_0\delta_{m,0}, \; -\infty<m<\infty, \quad (31)$$

$$-S_m\rho_m+e^{-ikh}C_{m-1}\tau_{m-1}=C_0\delta_{m,0}, \; -\infty<m<\infty. \quad (32)$$

{(31),(32)} represents an infinite system of linear equations in the unknowns $\tau_m$, $\rho_m$, $-\infty < m < \infty$. It will be noticed that each of these equations provides a relationship between FB reflection and transmission amplitudes, the orders $m$ of which differ by unity. For example, when $m=0$, Eq. (32) provides a relationship between the two main amplitudes of interest $\rho_0$ and $\tau_{-1}$. This fortunate "one-off" occurrence can be traced to the fact that the transmission coefficient $T(x)$ employed in Eqs. (19) and (20) has the functional form of a Fourier function $e^{2in\pi x/d}$ with $n=-1$ (see Eq. (7)). This can be utilized, together with the $\delta_{m,0}$ factor on the right-hand sides of (31) and (32), to halve the number of unknowns and the number of equations which must be solved. This can be seen by noting that (31) and (32) will be satisfied by $\tau_{m+1} = \rho_m = 0$ for all odd values of $m$, leaving only the $m$-even equations to be solved. These $m$-even equations of (31) and (32) may be written using both odd and even values of the integer $m$:

$$-C_{2m}\rho_{2m} + e^{ikh}S_{2m+1}\tau_{2m+1} = S_0\delta_{m,0}, \ -\infty < m < \infty, \quad (33)$$

$$-S_{2m}\rho_{2m} + e^{-ikh}C_{2m-1}\tau_{2m-1} = C_0\delta_{m,0}, \ -\infty < m < \infty. \quad (34)$$

A particular solution to {(33),(34)} alone is not necessarily a valid solution of the entire problem since it might not satisfy the global conservation of power. An "additional" solution must then be sought to supplement the already-found solution so that their sum would produce power conservation. This may be understood by realizing that the system {(33),(34)} is an "inhomogeneous" system of equations in that the free vector (the excitation-dependent right-hand-side of the system) is non-zero. In principle, the solution of this system should be the sum of a particular solution to the inhomogeneous system and the general solution to the homogeneous form of this system (i.e. the system {(33),(34)} with the free vector set to null) [47]. Since solutions to a homogeneous system involve an arbitrary factor, this factor can be adjusted to assure conservation of power. In what follows, the solutions to the inhomogeneous system {(33),(34)} will be denoted $\bar{\tau}_m$ and $\bar{\rho}_m$, and the solutions to the homogeneous form of these equations will be denoted $\tilde{\tau}_m$ and $\tilde{\rho}_m$, so that the total solution may be written

$$\rho_m = \bar{\rho}_m + \tilde{\rho}_m, \quad \tau_m = \bar{\tau}_m + \tilde{\tau}_m. \quad (35)$$

**a. Solution of Inhomogeneous Equations**

An accepted practice for solving the infinite linear system {(33),(34)} might be to truncate it so that $-N < m < N$, and to utilize standard methods for solving finite systems. The infinite system, however, is characterized by a bi-diagonal matrix the truncation of which can produce completely misleading results. For example, the truncation of Eq. (33) when $m=N$ would "truncate" the $\tau_{2N+1}$ term leaving $-C_{2N}\rho_{2N}=0$ which has the immediate solution $\rho_{2N}=0$. Using this solution in Eq. (34) when $m=N$ will lead to $\tau_{2N-1}=0$. Therefore, with successive substitutions, an erroneous solution for all unknowns can be found that is based solely on the truncation! An alternative method of solution must therefore be employed. The one presented below presumably would be applicable to any problem involving an infinite bi-diagonal system of linear equations.

To solve the inhomogeneous system {(33),(34)}, it is sufficient to arbitrarily set a "base" value of, say $\bar{\tau}_{2m'+1}$ for some particular value $m'$. Then, Eqs. (33) and (34) may be used successively to determine $\bar{\tau}_{2m+1}$, $\bar{\rho}_{2m}$ for every other value of $m$. We will choose a base value which will lead to a solution with special properties. We know from Eqs. (3) and (4) on which our model is based, that when the actual incidence angle $\psi_{inc}$ is the same as the designated incidence angle $\theta_{inc}$, the actual transmission angle $\psi_{trans}$ will be the same as the designated transmission angle $\theta_{trans}=0$. Furthermore, it was shown that under these conditions, there will be only two scattered waves [3],[10],[36],[37]: a specularly reflected wave (corresponding to the $\rho_0$ term in Eq. (16)) and an anomalously refracted wave (corresponding to the $\tau_{-1}$ term in Eq. (16)). The solution of {(33),(34)} that will provide this special two-wave property when $\psi_{trans}=0$ may be obtained by choosing the base value $\bar{\tau}_1 =0$. Successive substitution in Eqs. (33) and (34) reveals that this would result in

$$\bar{\rho}_{2m} = \bar{\tau}_{2m+1} = 0, \ m \geq 1. \quad (36)$$

Table 1 indicates the values of $\bar{\tau}_m$ and $\bar{\rho}_m$ for several other values of the index $m$. Their values for more negative values of $m$ than those in the table may again be obtained by successive application of Eqs. (33) and (34). The result is

$$\bar{\tau}_{2m-1} = e^{ikh}\frac{S_{2m}}{C_{2m-1}}\bar{\rho}_{2m}, \ \bar{\rho}_{2m} = e^{ikh}\frac{S_{2m+1}}{C_{2m}}\bar{\tau}_{2m+1}, \ m \leq -1. \quad (37)$$

It may be seen from Table 1 and Eq. (37) that all $\bar{\tau}_{2m+1}$, $\bar{\rho}_{2m}$, $m \leq -1$, contain the factor $S_{-1}$. But from (25) $S_{-1}=0$ when $\psi_{trans}=0$ (which occurs when $\psi_{inc}=\theta_{inc}$), so that in this case, $\bar{\tau}_{2m+1} = \bar{\rho}_{2m} = 0$ for $m \leq -1$. Combining this with Eq. (36) and Table 1 it may be concluded that when $\psi_{inc}=\theta_{inc}$, all the $\bar{\tau}_m$, $\bar{\rho}_m$ vanish except for $\bar{\tau}_{-1}$ and $\bar{\rho}_0$. Using $\tilde{\rho}_0 = \tilde{\tau}_{-1} = 0$ (which will be justified presently) in (35) along with the Table 1 expressions for $\bar{\tau}_{-1}$ and $\bar{\rho}_0$ yields

$$\rho_0 = \bar{\rho}_0 = -S_0/C_0, \quad \tau_{-1} = \bar{\tau}_{-1} = \frac{\gamma_0}{C_{-1}C_0}, \ \psi_{inc}=\theta_{inc}. \quad (38)$$

Table 1. Wave amplitude solutions to the inhomogeneous and homogeneous forms of Eqs. (33) and (34) derived from their successive evaluations, $m=2,1,0,-1,-2$ when $\bar{\tau}_1=0$. The complete solution is the sum of the inhomogeneous and homogeneous solutions (see Eq. (35)). The $\tilde{\tau}_{-1}$ and $\tilde{\rho}_{-2}$ have been set $\approx 0$ since they contain the factor $S_0\tilde{\rho}_0$ which includes the factor $S_0 S_{-1} \approx 0$ (see Eq. B1).

|  | Inhomogeneous solution |  | Homogeneous solution |
|---|---|---|---|
| $\rho_2=$ | $\bar{\rho}_2 = 0$ | + | $\tilde{\rho}_2 = \dfrac{1}{e^{2ikh}} \dfrac{C_1 C_0}{S_2 S_1} \tilde{\rho}_0$ |
| $\tau_1=$ | $\bar{\tau}_1 = 0$ | + | $\tilde{\tau}_1 = \dfrac{1}{e^{ikh}} \dfrac{C_0}{S_1} \tilde{\rho}_0$ |
| $\rho_0=$ | $\bar{\rho}_0 = -S_0 / C_0$ | + | $\tilde{\rho}_0$ |
| $\tau_{-1}=$ | $\bar{\tau}_{-1} = e^{ikh} \dfrac{\gamma_0}{C_{-1} C_0}$ | + | $\tilde{\tau}_{-1} = e^{ikh} \dfrac{S_0}{C_{-1}} \tilde{\rho}_0 \approx 0$ |
| $\rho_{-2}=$ | $\bar{\rho}_{-2} = e^{2ikh} \dfrac{S_{-1}\gamma_0}{C_{-2}C_{-1}C_0}$ | + | $\tilde{\rho}_{-2} = e^{2ikh} \dfrac{S_{-1}S_0}{C_{-2}C_{-1}} \tilde{\rho}_0 \approx 0$ |

Consistent with [3], [37] it will now be shown that the two-wave solution in Eq. (38) satisfies the power conservation equation which may be expressed as

$$\sum_n (\eta_n^\rho + \eta_n^\tau) = 1, \qquad (39)$$

where the sum is over all *propagating* FB wave components, and the $n$th mode power coupling coefficients for transmission and reflection are respectively

$$\eta_n^\tau = |\tau_n|^2 \gamma_n / \gamma_0, \quad \eta_n^\rho = |\rho_n|^2 \gamma_n / \gamma_0. \qquad (40)$$

Equation (39) expresses the fact that the energy emanating from the FP-HMS into the lower region $y<-h$ is the same as that entering the FP-HMS from the upper region $y>0$. Using Eqs. (38) and (40), Eq. (39) reduces to

$$|-S_0 / C_0|^2 \gamma_0 + \left|\dfrac{\gamma_0}{C_{-1}C_0}\right|^2 \gamma_{-1} = \gamma_0, \qquad (41)$$

which is an identity under the assumed $\psi_{inc}=\theta_{inc}$ conditions $C_{-1}=\gamma_{-1}=1$ given in Eq. (25). The two-wave solution in Eq. (38) therefore satisfies power conservation, thereby justifying the assumption $\tilde{\rho}_0 = \tilde{\tau}_{-1} = 0$ for this $\psi_{inc}=\theta_{inc}$ case.

Thus, the (inhomogeneous) solution for the $\bar{\tau}_{2m+1}$, $\bar{\rho}_{2m}$ given in Eqs. (36), (37) and Table 1 – which is valid for all values of $\psi_{inc}$ – has been shown to reduce to a two-wave power-conserving solution of {(33),(34)} for the designated angle of incidence $\psi_{inc}=\theta_{inc}$.

**b. Solution of Homogeneous Equations (Non-Designated Angle of Incidence)**

When power is not conserved by the solutions $\bar{\tau}_{2m+1}$, $\bar{\rho}_{2m}$ of the inhomogeneous system {(33),(34)}, these must be supplemented by solutions $\tilde{\tau}_{2m+1}, \tilde{\rho}_{2m}$ of the homogeneous form of Eqs. (33) and (34) (i.e. the same equations but with right-hand-side zero). That is, the total solution for the FB amplitudes will be given by Eq. (35). Just as in the case of the inhomogeneous system of equations, all amplitudes $\tilde{\tau}_{2m+1}$ and $\tilde{\rho}_{2m}$ can be expressed in terms of a "base" value which will be chosen as $\tilde{\rho}_0$. Then

$$\tilde{\tau}_1 = \dfrac{1}{e^{ikh}} \dfrac{C_0}{S_1} \tilde{\rho}_0, \qquad (42)$$

and successive application of Eqs. (33) and (34) with larger values of $m$ will lead to all the $\tilde{\tau}_{2m+1}, \tilde{\rho}_{2m}$, $m>0$, being expressed in terms of $\tilde{\rho}_0$. Similarly,

$$\tilde{\tau}_{-1} = e^{ikh} \dfrac{S_0}{C_{-1}} \tilde{\rho}_0, \qquad (43)$$

and successive application Eqs. (33) and (34) with smaller values of $m$ will lead to all the $\tilde{\tau}_{2m-1}, \tilde{\rho}_{2m}$, $m<0$, being expressed in terms of $\tilde{\rho}_0$. It is therefore seen that once an expression is available for $\tilde{\rho}_0$, expressions can be found for all the other $\tilde{\rho}_{2m}$ and $\tilde{\tau}_{2m+1}$. These are provided explicitly in Table 1.

**c. Solution for $\tilde{\rho}_0$**

The parameter $\tilde{\rho}_0$ will be chosen to satisfy the power conservation equation Eq. (39). Consider the total solution for $\rho_m$ and $\tau_m$ given in Table 1 by adding the solution of the homogeneous system to the solution of the inhomogeneous system. Using this in Eq. (40) and the result in Eq. (39) will produce an equation which can be solved for $|\tilde{\rho}_0|$. This is implemented in Appendix B for cases in which FB modes 0 and -1 are propagating for all values of $\psi_{inc}$ and mode 1 is propagating for small values of $\psi_{inc}$ (see Appendix C). It can, of course, be implemented in an entirely similar manner when other modes are propagating. Then

$$\tilde{\rho}_0 = |\tilde{\rho}_0| e^{i\varphi}, \qquad (44)$$

$$|\tilde{\rho}_0| \approx \begin{cases} \dfrac{S_{-1}}{C_{-1}}, \text{mode 1 not propagating} \\ \dfrac{S_{-1}}{C_{-1}} \dfrac{S_1}{C_1}, \text{mode 1 propagating} \end{cases} \qquad (45)$$

where Eq. (45) was obtained from Appendix B, and it was assumed that the $m=0$ and $m=-1$ FB modes are always propagating.

Although the magnitude of $\tilde{\rho}_0$ is provided by the power conservation condition, its phase $\varphi$ is not. However, as discussed in Appendix B, the values of the power coupling coefficients $\eta_n^\rho$ and $\eta_n^\tau$ are barely sensitive to $\varphi$.

## IV. RESULTS AND DISCUSSION

The closed form expressions given in Table 1 will now be tested by comparing them with full-wave computations of plane wave scattering from the FP-HMS, and with results for scattering from the ZT-HMS obtained using the methods of [37]. These comparisons will include both FB transmission / reflection amplitudes, and H-fields.

### 1. Two-Wave Solution

It was shown that when a plane wave is incident on the FP-HMS at an angle $\psi_{inc}=\theta_{inc}$, all the $\rho_m$ and $\tau_m$ in Eq. (16) will vanish except for $\rho_0$ and $\tau_{-1}$ which are given by Eq. (38). Using Eqs. (15), (16) and (38) in Eq. (14) will produce expressions for the z-directed H-fields above and below the FP-HMS that are consistent with standard HMS design theorems [3],[37],[36] :

$$H_>(x,y) = e^{i\alpha_0 x - i\beta_0 y} - \frac{S_0}{C_0}e^{i\alpha_0 x + i\beta_0 y}, \; H_<(x,y) = \frac{\gamma_0}{C_{-1}C_0}e^{i\alpha_{-1}x - i\beta_{-1}y} \quad (46)$$

where all elements are to be evaluated using $\psi_{inc}=\theta_{inc}$. The fields within the FP-HMS may be obtained from Eqs. (28) to (30), with $\sigma_{pa}$ and $\sigma_{pb}$ obtained from Eqs. (26) and (27):

$$\sigma_{pa} = C_0 e^{i\alpha_0 x_p} + S_0\rho_0 e^{i\alpha_0 x_p}, \; \sigma_{pb} = e^{ikh}S_{-1}\tau_{-1}e^{i\alpha_{-1}x} = 0. \quad (47)$$

These "closed-form" fields are displayed in Fig. 6(a) for $\theta_{inc}=80°$. Full-wave results computed by the CST program [48] applied to the structure of Fig. 5 appear in Fig. 6(b). In order to compare these FP-HMS fields with those of the abstract structure which it is meant to emulate, Fig. 6(c) displays the H-fields for the ZT-HMS [37]. Although significant specular reflection is apparent in this case (as expected from HMS theory [3],[10],[36],[37]), this extreme angle serves as a good case study for demonstrating our model.

It is clear that for both types of HMS in Fig. 6, the same interference pattern is apparent in the upper region, and the sole transmitted wave is consistent with the design parameter $\theta_{trans}=0$ in the lower region. It should be noted, too, that the excellent agreement between the closed form solution and the full wave (CST) solution includes the region of the FP-HMS itself! Furthermore, there is no evidence in Figs. 6(a) and 6(b) of the discrete nature of the FP-HMS, thereby demonstrating the homogenization capability of even electrically thick surfaces.

The conformity of Figs. 6(a) and 6(b) with Fig. 6(c) is also particularly noteworthy in that it contradicts the popular belief that a thin surface is required to produce metasurface phenomena. It demonstrates the non-intuitive result that, despite its electrically large thickness, the proposed FP-HMS device is able to *meticulously and accurately* emulate the abstract zero-thickness HMS response.

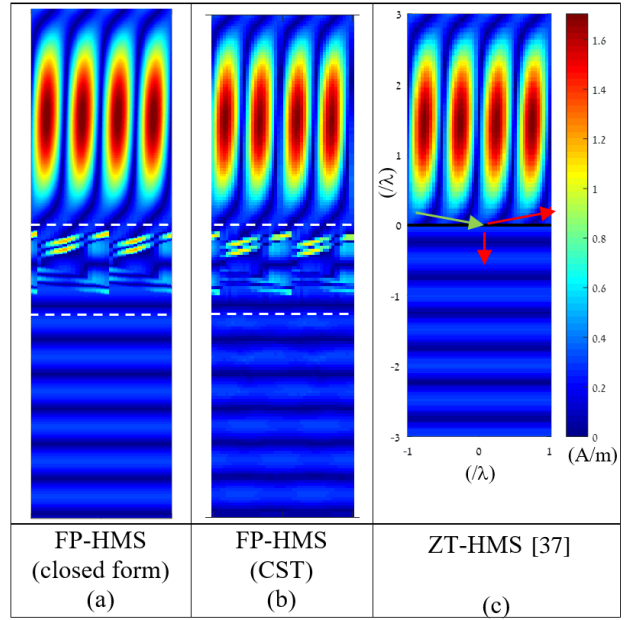

| FP-HMS (closed form) | FP-HMS (CST) | ZT-HMS [37] |
|---|---|---|
| (a) | (b) | (c) |

FIG. 6. Color image of $|Re(H)|$ (a) for the closed-form solution for the FP-HMS given in Eqs. (30) and (46), (b) for the FP-HMS as computed by CST, and (c) for the ZT-HMS as computed by Matlab [37], for an incident field $|H_{inc}| = 1$ A/m. In each case, $\psi_{inc}=\theta_{inc}=80°$, $\theta_{trans}=0$. The region extends $2d$ in the x-direction, and $3\lambda$ above the upper surface and below the lower surface. The arrows in (c) indicate the incidence, reflection and transmission directions. The images in (a) and (b) are longer than that in (c) because of the finite thickness $h=1.3\lambda$ of the FP-HMS.

### 2. General FP-HMS Solution: Compatibility with Full-Wave Solution

The results in the previous sub-section were obtained for incidence angle $\psi_{inc}=\theta_{inc}$. Results will now be considered for other values of $\psi_{inc}$. Full-wave solutions obtained by applying the CST computer program to the FP-HMS structure in Fig. 5 will be compared with the closed-form solutions given in Table 1 supplemented by Eqs. (44) and (45).

For design parameters $\theta_{inc}=80°$, $\theta_{trans}=0$, Fig. 7 provides the values of the specular reflection power-coupling efficiency $\eta_0^\rho$ and the anomalous transmission power coupling efficiency $\eta_{-1}^\tau$ as functions of the incidence angle $\psi_{inc}$ for both the closed-form formulas and the full-wave

CST-calculated solution. The agreement between these two solutions is excellent over the *entire* range of $\psi_{inc}$ including the small values of $\psi_{inc}$ near which the slope of the $\eta_0^\rho$ curve is discontinuous.

Fig. 7 also includes results based on the Reference [37] solution to the ZT-HMS which indicate good agreement with the FP-HMS results for $\eta_{-1}^\tau$ over the entire range of $\psi_{inc}$, and good agreement for $\eta_0^\rho$ for larger values of $\psi_{inc}$. However, this good agreement ceases for smaller values of $\psi_{inc}$. This discrepancy can be traced to the fact that in [37] the existence of a solution to the "homogeneous" form of {(33),(34)} was overlooked, which is tantamount to $\tilde{\rho}_0 = 0$. Since from Table 1 this does not affect the expression for $\tau_{-1}$ (since $\tilde{\tau}_{-1} \approx 0$ anyway), the Ref. [37] ZT-HMS results for the dominant anomalous refraction efficiency $\eta_{-1}^\tau$ are seen in Fig. 7 to agree with the other results.

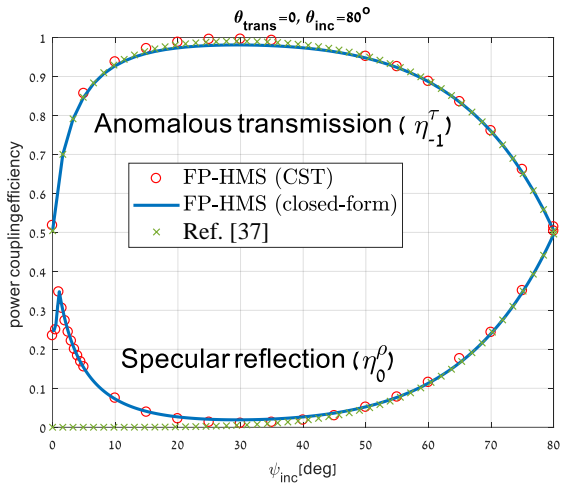

FIG. 7. Power coupling efficiency comparisons between CST full-wave solutions applied to the FP-HMS shown in Fig. 5; closed-form solutions for $\rho_0$ and $\tau_{-1}$ in Table 1 obtained with the aid of Eqs. (44) and (B17); and ZT-HMS solutions [37]. The FP-HMS height $h=1.3\lambda$, the design transmission angle $\theta_{trans}=0$, and the design incidence angle $\theta_{inc}=80°$. The $\psi_{inc}=80°$ results correspond to those in Fig. 6.

Fig.7 considers power coupling efficiencies for a single designated incidence angle $\theta_{inc}=80°$ over the domain of actual incidence angles $0<\psi_{inc}<80°$. Fig. 8 provides such comparisons for *other* designs (other values of $\theta_{inc}$) over a domain of $\psi_{inc}$ within which the results can be easily distinguished from each other. The power coupling efficiencies $\eta_{-1}^\tau$, $\eta_0^\rho$, and $\eta_1^\tau$ are provided in Figs. 8(a), 8(b) and 8(c), respectively. It should be noted that the $\tau_1$ FB wave only propagates for small values of $\psi_{inc}$. The maximum value of $\psi_{inc}$ for which FB mode 1 propagates depends on $\theta_{inc}$ as given in Table C-1 in Appendix C. This is the reason that the $\psi_{inc}$-extents of the $\eta_1^\tau$ curves in Fig. 8(c) differ. These values of $\psi_{inc}$ are also the crossover points between the positive and negative slopes of the curves in Fig. 8(b). The excellent agreement between the closed-form and full-wave solutions in all parts of Fig. 8 is indicative of the success of the closed-form solution for predicting the fields scattered from the FP-HMS, and of the design of Fig. 5 for producing desired HMS scattering effects.

The results above related to the amplitudes of the *propagating* FB field components. Fields near the boundaries of the FP-HMS are affected by the *evanescent* FB components as well. The closed-form amplitudes of $\rho_n, \tau_n$, $-N \leq n \leq N$, $N=15$ – which include *both* propagating waves (light-colored bars) and evanescent waves (dark-colored bars) – are presented in the first column of Fig. 9 for several values of $\psi_{inc}$ for the case $\theta_{inc}=80°$ as in Fig. 7. It may be seen that $|\rho_{2m}|$, $|\tau_{2m+1}|$ decrease with $|m|$.

Also shown in Fig. 9 are images for the fields based on the closed-form amplitudes, and for the full-wave fields computed by CST (Fig. 6 contains such a comparison of field images for $\psi_{inc}=80°$). Unlike the power coupling efficiencies, the "closed-form" fields are dependent on the phase of $\tilde{\rho}_0$ which has not been determined by the formulation, but which can be empirically deduced from the FB amplitude magnitudes and the CST field snapshots.

The excellent agreement between the closed-form predictions and the full-wave solutions validates the FP-HMS design for producing the desired anomalous refraction, and demonstrates the consistency of the closed-form expressions for predicting field effects. The capability for obtaining excellent agreement for the fields shown in Fig. 9 for the designated incidence angle $\theta_{inc}=80°$ was observed as well for smaller values of $\theta_{inc}$. While it is irksome that the mathematical derivation did not inherently provide the phase $\varphi$ of the added homogeneous solution (Eq. (44)), it is also interesting if not fascinating, and a matter for further investigation.

Furthermore, frequency-response results displayed in Appendix E indicate that for common commercially available low-loss microwave substrates, the coupling efficiency of the inhomogeneous metasurface is only slightly affected, remaining above 80% for a fractional bandwidth of ~4%. This is comparable to previously reported HMS designs for anomalous refraction [1], indicating the practical viability of the proposed FP-HMS configuration.

Before concluding, we would like to stress that the availability of the solution in closed form provides the possibility of performing analyses which would otherwise be quite tedious. For example, the angle of incidence $\psi_{inc}$ can easily be found for which the power coupling efficiency $\eta_{-1}^\tau$ is optimum for anomalous refraction. This "optimum" $\psi_{inc}$ will generally differ from the designated incidence angle $\theta_{inc}$, since the latter was defined to provide a two-wave solution, *not* to optimize the transmission efficiency.

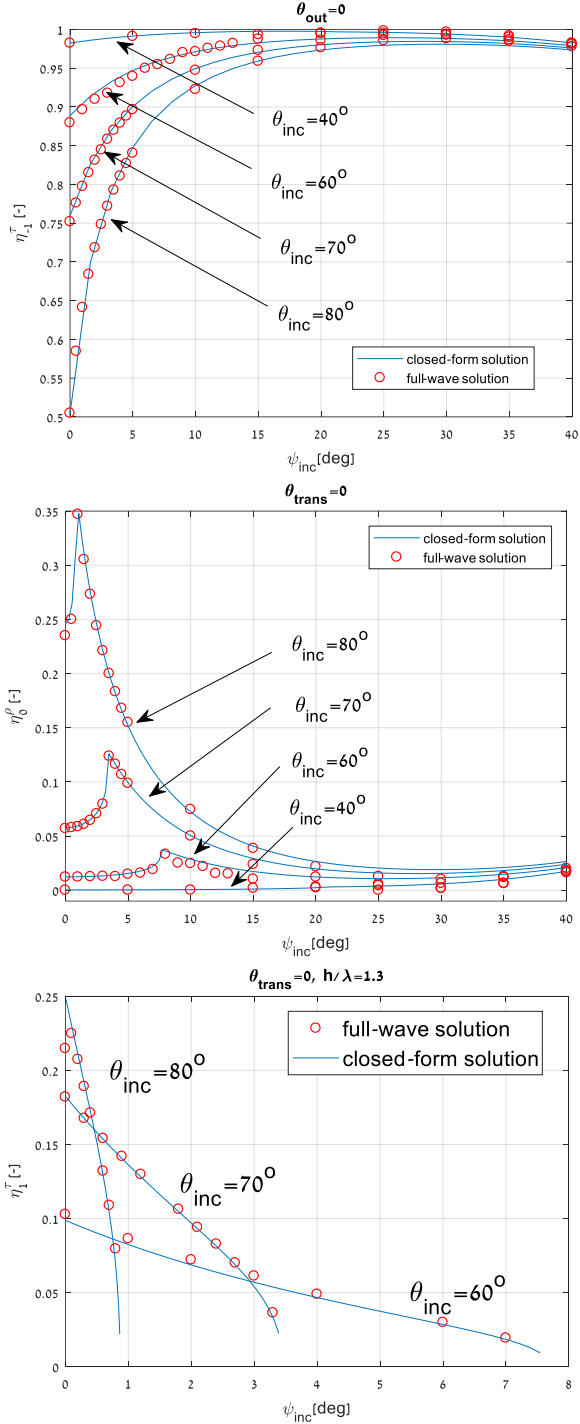
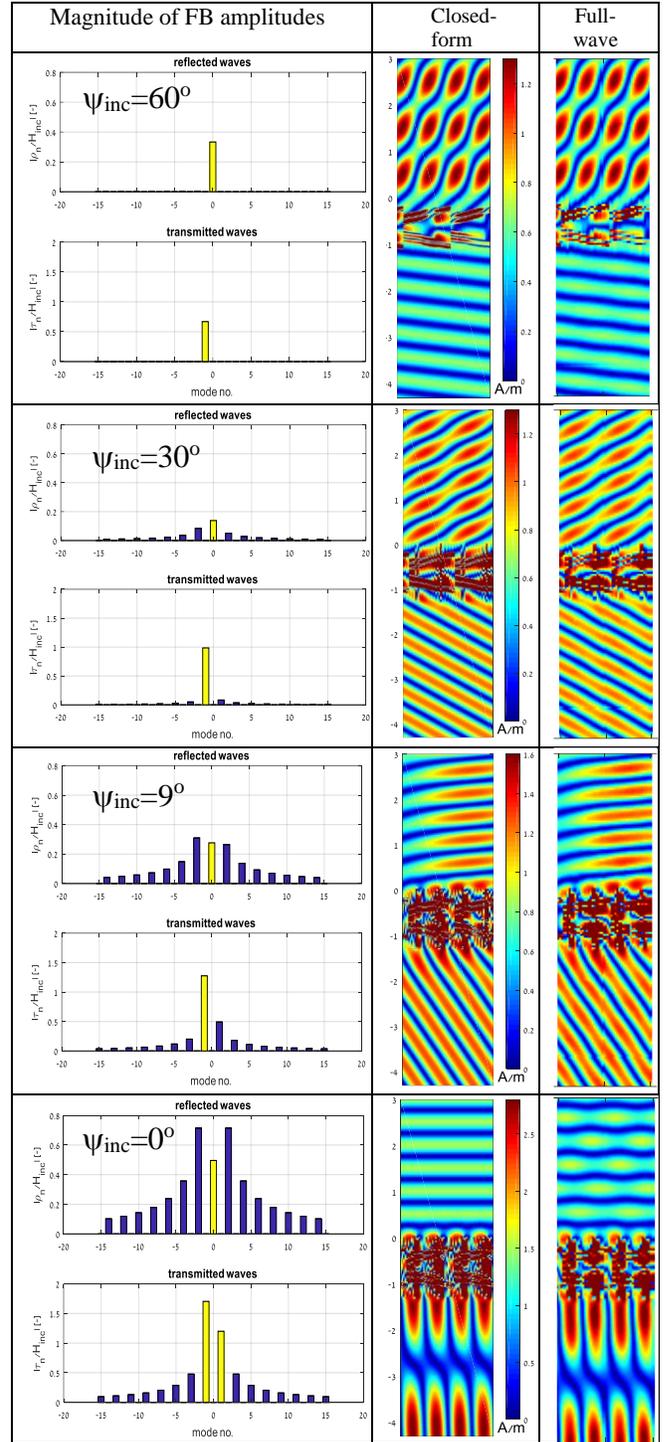

FIG. 8. Power coupling efficiency comparisons between CST full-wave solutions applied to the FP-HMS shown in Fig. 5 and closed-form solutions. The FP-HMS height $h=1.3\lambda$, the design transmission angle $\theta_{trans}=0$, and the design incidence angles $\theta_{inc}=40, 60, 70$ and $80°$. (a) $\eta^\tau_{-1}$; (b) $\eta^\rho_0$; (c) $\eta^\tau_1$. In each case, closed-form expressions were obtained from Table 1 with the aid of Eqs. (44) and (45).

FIG. 9. Magnitudes of closed-form FB amplitudes, $|Re(H)|$ based on these amplitudes, and $|Re(H)|$ based on full-wave CST-computed fields for the FP-HMS designed using $\theta_{inc}=80°$, $\theta_{trans}=0$, $h=1.3\lambda$ for incidence angles $\psi_{inc}=0, 9, 30$ and $60°$, and incident field $|H_{inc}|=1$ A/m. Distances are normalized to wavelength. Light-colored bars in column 1 represent propagating waves, dark colored bars represent evanescent waves.

From Table 1, the FB amplitude of interest is given by

$$\tau_{-1} = e^{ikh}\left(\frac{\gamma_0}{C_0 C_{-1}}\right), \qquad (48)$$

so that the power coupling efficiency is

$$\eta_{-1}^{\tau} = |\tau_{-1}|^2 \gamma_{-1}/\gamma_0 = F(C_0)F(C_{-1}), \quad F(C_n) \equiv \frac{2C_n - 1}{C_n^2}, \quad (49)$$

where, from (23) and (24), $C_0$ depends only on $\psi_{inc}$ and $C_{-1}$ depends only on $\psi_{trans}$ (which also depends implicitly on $\psi_{inc}$). The value of $\psi_{inc}$ for which $\eta_{-1}^{\tau}$ is extremum is determined by solving $d\eta_{-1}^{\tau}/d\psi_{inc}=0$, and is found to satisfy

$$\tan^2\left(\frac{\psi_{trans}}{2}\right)\frac{\sin\psi_{trans}}{\cos^2\psi_{trans}} = -\tan^2\left(\frac{\psi_{inc}}{2}\right)\frac{\sin\psi_{inc}}{\cos^2\psi_{inc}}. \quad (50)$$

This implies that the incidence angle $\psi_{inc}$ at which $\eta_{-1}^{\tau}$ is largest for a given FP-HMS design is that for which the transmission angle $\psi_{trans}$ is the negative of $\psi_{inc}$ (where the impedance mismatch between incident and refracted waves vanishes). For $\theta_{inc}=80°$, $\theta_{trans}=0$, this corresponds to $\psi_{inc}=-\psi_{trans}\approx30°$, consistent with the closed-form results shown in Fig. 7. The fields image for this case is shown in the $\psi_{inc}=30°$ results of Fig. 9, where it is clear that $\psi_{trans}\approx-\psi_{inc}$.

## V. CONCLUSION

The design of a Fabry-Pérot Huygens' metasurface has been proposed for implementation of Huygens' metasurfaces. It was shown to produce the same type of anomalous refraction as that provided by an "abstract" zero-thickness HMS characterized by a surface impedance and admittance. An analytical method has been derived to provide closed-form solutions for the FB spectrum of waves scattered from this FP-HMS, offering phenomenological insights into the propagation processes that were heretofore unavailable. Full agreement has been shown between this closed-form solution and the full-wave solution provided by CST. Agreement was also found between these FP-HMS solutions and solutions to the ZT-HMS provided in [37] for incident angles of greatest interest, and the reasons for discrepancies at other incidence angles are well-understood. This validates the proposed FP-HMS design as a homogenized metasurface, and the validity of the surface-homogenization approximation for appropriately designed electrically thick structures. In addition, the newly proposed structure, which can be designed semi-analytically up to the detailed physical layout, could assist in practical realizations of HMS-based devices, a nontrivial task to date. Finally, since the Fabry-Pérot mechanism is universal and only requires media with differing wave velocity, the FP-HMS concept could be very useful for other wave phenomena (e.g., the transmissive counterpart of [35] for acoustic propagation).

## APPENDIX

### A. Boundary Conditions

With the aid of Eqs. (14) to (18), the conditions for the continuity of the E and H field components tangent to the upper ($y=0$) and lower ($y=-h$) boundaries may be written:

$$\left[H_{inc}(x,y) + H_{ref}(x,y) = \sigma_{pa}e^{-iky} + T_p\sigma_{pb}e^{iky}\right]_{y=0}, \quad (A1)$$

$$\left[\frac{\partial H_{inc}(x,y)}{\partial y} + \frac{\partial H_{ref}(x,y)}{\partial y} = -ik\sigma_{pa}e^{-iky} + ikT_p\sigma_{pb}e^{iky}\right]_{y=0}, \quad (A2)$$

$$\left[H_{trans}(x,y) = T_p\sigma_{pa}e^{-iky} + \sigma_{pb}e^{iky}\right]_{y=-h}, \quad (A3)$$

$$\left[\frac{\partial H_{trans}(x,y)}{\partial y} = -ikT_p\sigma_{pa}e^{-iky} + ik\sigma_{pb}e^{iky}\right]_{y=-h}. \quad (A4)$$

These four equations can be appreciably simplified by solving for $\sigma_{pa}$ and $\sigma_{pb}$. From Eqs. (A1) and (A2), it is easily found that

$$\sigma_{pa} = \frac{1}{2}\left[H_{inc}(x,y) + H_{ref}(x,y) - \frac{1}{ik}\frac{\partial H_{inc}(x,y)}{\partial y} - \frac{1}{ik}\frac{\partial H_{ref}(x,y)}{\partial y}\right]_{y=0}. \quad (A5)$$

Similarly, from Eqs. (A3) and (A4),

$$\sigma_{pb} = \frac{e^{ikh}}{2}\left[H_{trans}(x,y) + \frac{1}{ik}\frac{\partial H_{trans}(x,y)}{\partial y}\right]_{y=-h}. \quad (A6)$$

Substituting Eqs. (A5) and (A6) in Eqs. (A1) and (A3) produces

$$-\frac{1}{2}\left[H_{ref}(x,y) + \frac{1}{ik}\frac{\partial H_{ref}(x,y)}{\partial y}\right]_{y=0}$$
$$+T_p\frac{e^{ikh}}{2}\left[H_{trans}(x,y) + \frac{1}{ik}\frac{\partial H_{trans}(x,y)}{\partial y}\right]_{y=-h}$$
$$=\frac{1}{2}\left[H_{inc}(x,y) + \frac{1}{ik}\frac{\partial H_{inc}(x,y)}{\partial y}\right]_{y=0}, \quad (A7)$$

$$T_p e^{ikh} \frac{1}{2}\left[H_{ref}(x,y) - \frac{1}{ik}\frac{\partial H_{ref}(x,y)}{\partial y}\right]_{y=0}$$
$$+\frac{1}{2}\left[H_{trans}(x,y) + \frac{1}{ik}\frac{\partial H_{trans}(x,y)}{\partial y}\right]_{y=-h} - \left[H_{trans}(x,y)\right]_{y=-h}$$
$$= -T_p e^{ikh}\frac{1}{2}\left[H_{inc}(x,y) - \frac{1}{ik}\frac{\partial H_{inc}(x,y)}{\partial y}\right]_{y=0}. \quad (A8)$$

Finally, utilizing Eqs. (15) and (16), carrying out the differentiations, and approximating $T_p \approx e^{-i2\pi x/d}$, Eqs. (A7) and (A8) can be written

$$-\sum_{n=-\infty}^{\infty} \rho_n C_n e^{i\left(\frac{2n\pi}{d}\right)x} + e^{-i2\pi x/d} e^{ikh} \sum_{n=-\infty}^{\infty} \tau_n S_n e^{i\left(\frac{2n\pi}{d}\right)x} = S_0, \quad (A9)$$

$$-\sum_{n=-\infty}^{\infty} \rho_n S_n e^{i\left(\frac{2n\pi}{d}\right)x} + e^{i2\pi x/d} e^{-ikh} \sum_{n=-\infty}^{\infty} \tau_n C_n e^{i\left(\frac{2n\pi}{d}\right)x} = C_0, \quad (A10)$$

where

$$S_n = \frac{1}{2}(1-\gamma_n), \quad C_n = \frac{1}{2}(1+\gamma_n), \quad \gamma_n = \frac{\beta_n}{k}. \quad (A11)$$

### B. Determination of $\tilde{\rho}_0$

The expression for $|\tilde{\rho}_0|$ is not known, but may be found from the conservation of power as expressed in Eq. (39), and expanded as

$$|\rho_0|^2 \gamma_0 + |\tau_{-1}|^2 \gamma_{-1} + |\tau_1|^2 \gamma_1 = \gamma_0. \quad (B12)$$

This equation involves the moduli of $\tau_{2n+1}$ and $\rho_{2n}$ terms appearing in Table 1. For any term that is proportional to $\tilde{\rho}_0$ (as is the case for $\tau_1$ in Table 1), the modulus of that term is not dependent on the phase of $\tilde{\rho}_0$. On the other hand, any term that is not proportional to $\tilde{\rho}_0$ (as in $\rho_0$ and $\tau_{-1}$ in Table 1), its modulus would be dependent on the phase of $\tilde{\rho}_0$. For example, $|\tau_1| = C_0 |\tilde{\rho}_0|/S_1$ which does not depend on the phase of $\tilde{\rho}_0$; but $|\tau_{-1}| = |C_0 + \bar{\rho}_0 S_0 + S_0 \tilde{\rho}_0|/C_{-1}$ which does depend on the phase of $\tilde{\rho}_0$. From Table 1, the only cases in which $\tau_n$ and $\rho_n$ are not proportional to $\tilde{\rho}_0$ are $\rho_0$ and $\tau'_{-1}$, and both involve $\tilde{\rho}_0$. It may therefore be concluded that the solution of Eq. (B12) for $|\tilde{\rho}_0|$ would depend on the phase of $\tilde{\rho}_0$. It is shown below that this dependence of $|\tilde{\rho}_0|$ on the phase of $\tilde{\rho}_0$ is very weak and can be ignored.

Substituting the expressions of Table 1 in Eq. (B12),

$$\left|-\frac{S_0}{C_0} + \tilde{\rho}_0\right|^2 \gamma_0 + \left|\frac{\gamma_0}{C_0 C_{-1}} + \frac{S_0}{C_{-1}}\tilde{\rho}_0\right|^2 \gamma_{-1} + \left|\frac{C_0}{S_1}\tilde{\rho}_0\right|^2 \gamma_1 = \gamma_0, \quad (B13)$$

where $|e^{ikh}|=1$ was used. As explained in Appendix C, the last term on the left (containing the factor $\gamma_1$) will only be present if the $\rho_1$, $\tau_1$ FB components are propagating (i.e. $\gamma_1$ is real). In order to account for the phase of $\tilde{\rho}_0$, substitute

$$\tilde{\rho}_0 = |\tilde{\rho}_0| e^{i\varphi} = |\tilde{\rho}_0|\cos\varphi + i|\tilde{\rho}_0|\sin\varphi, \quad (B14)$$

where $\varphi$ is real, and collect terms:

$$\left(\gamma_0 + \frac{\gamma_{-1}S_0^2}{C_{-1}^2} + \frac{C_0^2}{S_1^2}\gamma_1\right)|\tilde{\rho}_0|^2 + 2\frac{S_0}{C_0}\gamma_0\cos\varphi\left(\frac{\gamma_{-1}}{C_{-1}^2}-1\right)|\tilde{\rho}_0|$$
$$+\gamma_0\left(\frac{S_0^2}{C_0^2} + \frac{\gamma_0\gamma_{-1}}{C_0^2 C_{-1}^2} - 1\right) = 0 \quad (B15)$$

Finally, using the identities of Eq. (22),

$$\left(1 + \frac{\gamma_{-1}S_0^2}{\gamma_0 C_{-1}^2} + \frac{\gamma_1 C_0^2}{\gamma_0 S_1^2}\right)|\tilde{\rho}_0|^2 - 2\cos\varphi\frac{(S_0 S_{-1})S_{-1}}{C_0 C_{-1}^2}|\tilde{\rho}_0| - \frac{\gamma_0 S_{-1}^2}{C_0^2 C_{-1}^2} = 0, \quad (B16)$$

which can be solved for $|\tilde{\rho}_0|$ using the quadratic equation formula. Since $S_0 S_{-1} \ll 1$ (see Appendix D), the first order term can be ignored, and the solution can be approximated as

$$|\tilde{\rho}_0|^2 \approx \frac{\frac{\gamma_0 S_{-1}^2}{C_0^2 C_{-1}^2}}{1 + \frac{\gamma_{-1}S_0^2}{\gamma_0 C_{-1}^2} + \frac{\gamma_1 C_0^2}{\gamma_0 S_1^2}} \approx \frac{\gamma_0^2 S_{-1}^2}{C_0^2 C_{-1}^2 \gamma_0 + \gamma_{-1}C_0^2 S_0^2 + \frac{\gamma_1 C_0^4 C_{-1}^2}{S_1^2}},$$
(B17)

or

$$|\tilde{\rho}_0| \approx \frac{S_{-1}}{C_{-1}}\frac{1}{\left(\frac{\gamma_1}{S_1^2}+1\right)^{1/2}} = \frac{S_{-1}}{C_{-1}}\frac{S_1}{C_1} \quad (B18)$$

which does not depend on the phase $\varphi$. Therefore, the magnitude of $\tilde{\rho}_0$ is only weakly dependent on its phase. If mode 1 is not propagating, this reduces to

$$|\tilde{\rho}_0| \approx \frac{S_{-1}}{C_{-1}}, \quad (B19)$$

where the identities in Eq. (22) were utilized along with Eq. (D1).

Not only is the modulus of $\tilde{\rho}_0$ not sensitive to the phase $\varphi$, but also the values of the power coupling coefficients $\eta_n^\rho$ and $\eta_n^\tau$ are not sensitive to $\varphi$. This may be seen from their definition in (40) which implies, as mentioned above, that if $\rho_n$ (or $\tau_n$) can be written as only a single term $\bar{\rho}_n$ or $\tilde{\rho}_n$ (or $\bar{\tau}_n$ or $\tilde{\tau}_n$), then $|\rho_n|^2$ (or $|\tau_n|^2$) would be independent of the phase of $\tilde{\rho}_0$. Referring to Table 1, this is seen to be the case for all FB amplitudes except $\rho_0$ which *is* composed of the sum of two terms, $\bar{\rho}_0$ and $\tilde{\rho}_0$. However, for large values of $\psi_{inc}$ $\bar{\rho}_0$ is dominant and for small values of $\psi_{inc}$ $\tilde{\rho}_0$ is dominant; in either case $|\rho_0|^2$ is not sensitive to the phase of $\tilde{\rho}_0$. In the interim region, both $\bar{\rho}_0$ and $\tilde{\rho}_0$ are small so that the phase of $\tilde{\rho}_0$ only affects the degree of smallness; for the case $\theta_{inc}=80°$, $\theta_{trans}=0$, the interim region is near $\psi_{inc}=30°$ and the largest difference in $|\rho_0|^2$ which can be caused by the phase of $\tilde{\rho}_0$ is 0.025.

### C. Maximum $\psi_{inc}$ for Mode 1 Propagation

The HMS field consists of a sum of FB plane waves, but only several of these waves are propagating; the remainder are evanescent. The *n*th mode is propagating if $\beta_n$ in Eq. (17) is real; otherwise it is evanescent. Therefore, from Eqs. (17) and (5), mode $n$ will propagate if

$$1 > |\sin\psi_{inc} + n|\sin\theta_{trans} - \sin\theta_{inc}\|. \quad (C1)$$

Since we are generally interested in a design for which $\theta_{trans}=0$, the condition for mode 1 to propagate is

$$\sin\psi_{inc} < 1 - \sin\theta_{inc}. \quad (C2)$$

The maximum values of $\psi_{inc}$ for which the $n=1$ mode will propagate are given in Table C-1 for several values of $\theta_{inc}$.

Table C-1. Maximum values of $\psi_{inc}$ as a function of $\theta_{inc}$ for propagation of mode 1.

| $\theta_{inc}$ (°) | $\psi_{inc}$ (°) |
|---|---|
| 80 | 0.87 |
| 70 | 3.46 |
| 60 | 7.70 |
| 50 | 13.53 |
| 40 | 20.93 |

### D. The small quantity $S_0 S_{-1}$

Although the expressions for the FB amplitudes are given in closed form, many of them can be simplified further by noting that the quantity $S_0 S_{-1}$ is extremely small,

$$S_0 S_{-1} \ll 1, \quad (D1)$$

so that terms in which it appears as a factor can be ignored relative to other terms of order unity. Fig. D1 displays its magnitude as a function of the incidence angle $\psi_{inc}$ for several values of designated incidence angle $\theta_{inc}$ when the designated transmission angle is $\theta_{trans}=0$.

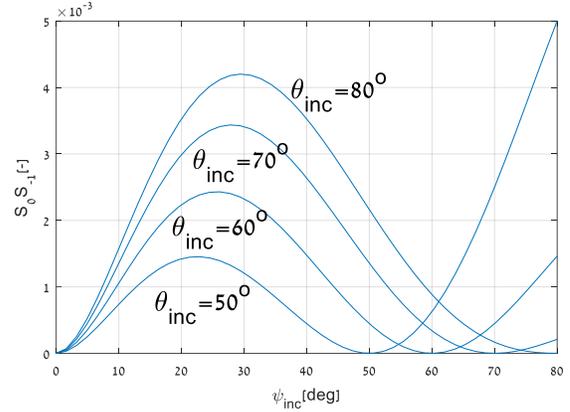

FIG. D1. The smallness of $S_0 S_{-1}$ as a function of incidence angle $\psi_{inc}$ for different values of designated incidence angle $\theta_{inc}$ ($\theta_{trans}=0$).

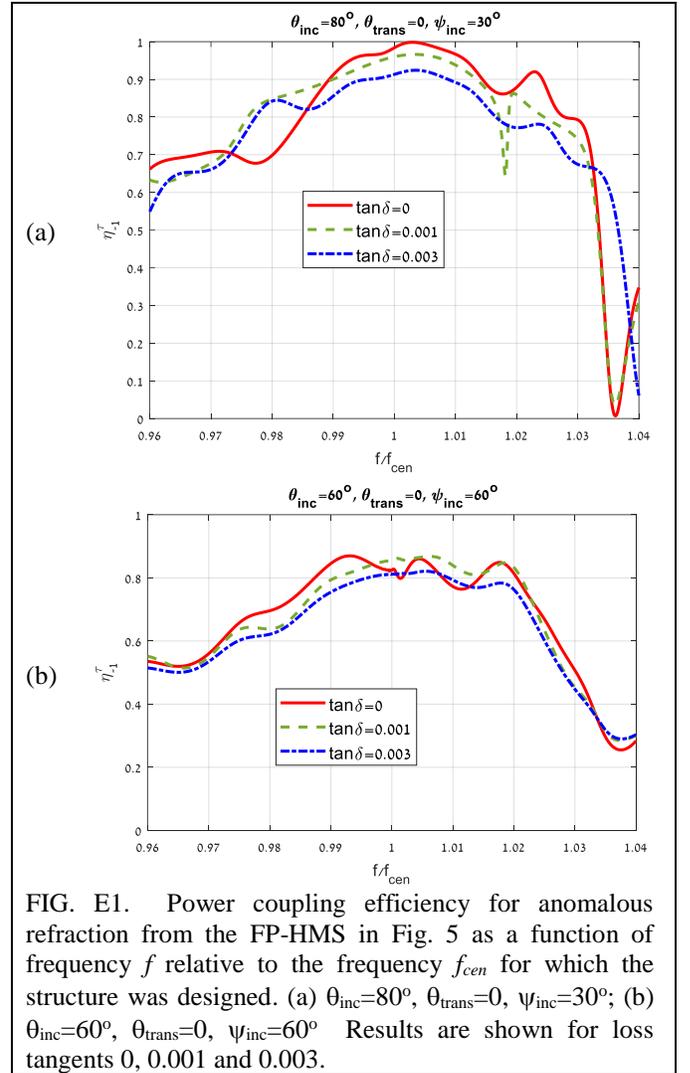

FIG. E1. Power coupling efficiency for anomalous refraction from the FP-HMS in Fig. 5 as a function of frequency $f$ relative to the frequency $f_{cen}$ for which the structure was designed. (a) $\theta_{inc}=80°$, $\theta_{trans}=0$, $\psi_{inc}=30°$; (b) $\theta_{inc}=60°$, $\theta_{trans}=0$, $\psi_{inc}=60°$ Results are shown for loss tangents 0, 0.001 and 0.003.

## E. Frequency Response Examples

The full-wave-simulated frequency response for two arbitrarily-chosen sets of $\theta_{inc}$ and $\psi_{inc}$ are shown in Fig. E1 for several realistic loss tangents. It can be observed that for the two different cases, with practical dielectric losses, high efficiency above 80% is retained for a moderate bandwidth of 4%, thereby indicating the practical applicability of the proposed concept.

______________________________

[1] C. Pfeiffer and A. Grbic, "Metamaterial Huygens' surfaces: tailoring wave fronts with reflectionless sheets," Phys. Rev. Lett. **110**, p. 197401 (2013).

[2] F. Monticone, N. M. Estakhri, and A. Alù, "Full control of nanoscale optical transmission with a composite metascreen," Phys. Rev. Lett. **110**, p. 203903 (2013).

[3] M. Selvanayagam and G. V. Eleftheriades, "Discontinuous electromagnetic fields using orthogonal electric and magnetic currents for wavefront manipulation," Opt. Express **21**, pp. 14409-14429 (2013).

[4] E. F. Kuester, M. A. Mohamed, M. Piket-May, and C. L. Holloway, "Averaged transition conditions for electromagnetic fields at a metafilm," IEEE Trans. Antennas Propag., **51,** pp. 2641–2651 (2003).

[5] C. L. Holloway, A. Dienstfrey, E. F. Kuester, J. F. O'Hara, A. K. Azad, and A. J. Taylor, "A discussion on the interpretation and characterization of metafilms/metasurfaces: The two-dimensional equivalent of metamaterials," Metamaterials, **3**, pp. 100–112 (2009).

[6] C. L. Holloway, E. F. Kuester, and A. Dienstfrey. "Characterizing metasurfaces/metafilms: The connection between surface susceptibilities and effective material properties," IEEE Antennas Wireless Propag. Lett., **10**, pp. 1507–1511 (2011).

[7] C. L. Holloway, E. F. Kuester, J. A. Gordon, J. O'Hara, J. Booth, and D. R. Smith. "An overview of the theory and applications of metasurfaces: The two-dimensional equivalents of metamaterials." IEEE Antennas Propag. Mag., **54,** pp**.** 10–35 (2012).

[8] S. A. Tretyakov, "Metasurfaces for general transformations of electromagnetic fields," Proc. R. Soc. A., **373**, p. 20140362 (2015).

[9] K. Achouri, M. A. Salem, and C. Caloz, "General metasurface synthesis based on susceptibility tensors IEEE Trans. Antennas Propag., **63**, pp. 2977–2991 ( 2015).

[10] A. Epstein and G. V. Eleftheriades, "Huygens' metasurfaces via the equivalence principle: design and applications," J. Opt. Soc. Am. B, **33**, 2, p. A31 (2016).

[11] N. M. Estakhri and A. Alù, "Recent progress in gradient metasurfaces," J. Opt. Soc. Am. B, **33**, p. A21 (2016).

[12] C. Pfeiffer and A. Grbic, "Millimeter-Wave Transmitarrays for Wavefront and Polarization Control," IEEE Trans. Microw. Theory Tech., **61**, pp. 4407–4417 (2013).

[13] C. Pfeiffer and A. Grbic, "Bianisotropic Metasurfaces for Optimal Polarization Control: Analysis and Synthesis," Phys. Rev. Appl., **2**, p. 044011 (2014).

[14] J. P. S. Wong, M. Selvanayagam, and G. V. Eleftheriades, "Design of unit cells and demonstration of methods for synthesizing Huygens metasurfaces," Photonics Nanostructures - Fundam. Appl., **12**, pp. 360–375 (2014).

[15] J. P. S. Wong, M. Selvanayagam, and G. V Eleftheriades, "Polarization considerations for scalar Huygens metasurfaces and characterization for 2-D refraction," IEEE Trans. Microw. Theory Tech., **63**, pp. 913–924 (2015).

[16] A. Epstein, J. P. S. Wong, and G. V. Eleftheriades, "Cavity-excited Huygens' metasurface antennas for near-unity aperture illumination efficiency from arbitrarily large apertures," Nat. Commun., **7**, p. 10360 (2016).

[17] M. Chen, E. Abdo-Sánchez, A. Epstein, and G. V. Eleftheriades, "Theory, design, and experimental verification of a reflectionless bianisotropic Huygens' metasurface for wide-angle refraction," Phys. Rev. B **97**, 12, p. 125433 (2018).

[18] G. Lavigne, K. Achouri, V. S. Asadchy, S. A. Tretyakov, and C. Caloz, "Susceptibility derivation and experimental demonstration of refracting metasurfaces without spurious diffraction," IEEE Trans. Antennas Propag., **66**, pp. 1321-1330 (2018).

[19] D. Kwon, G. Ptitcyn, A. Díaz-Rubio, S. A. Tretyakov, "Transmission magnitude and phase control for polarization-preserving reflectionless metasurfaces," Phys. Rev. Appl., **9**, pp. 034005-1-034005-12 (2018).

[20] F. S. Cuesta, I. A. Faniayeu, V. S. Asadchy, S. A. Tretyakov, " Planar broadband Huygens' metasurfaces for wave manipulations," IEEE Trans. Antennas Propag., **66**, pp. 7117-7127 (2018).

[21] A. Yariv, P. Yeh, "Photonics – Optical Electronics in Modern Communications", Oxford University Press (2007).

[22] Philippe Lalanne, Simion Astilean, Pierre Chavel, Edmond Cambril, and Huguette Launois, "Blazed binary subwavelength gratings with efficiencies larger than those of conventional échelette gratings," Opt. Lett. **23**, 1081-1083 (1998).

[23] Z. Bomzon, G. Biener, V. Kleiner, and E. Hasman, "Radially and azimuthally polarized beams generated by space-variant dielectric subwavelength gratings," Opt. Lett., **27**, p. 285 (2002).

[24] E. Hasman, V. Kleiner, G. Biener, and A. Niv, "Polarization dependent focusing lens by use of quantized Pancharatnam-Berry phase diffractive optics," Appl. Phys. Lett., **82**, pp. 328–330 (2003).

[25] D. Lin, P. Fan, E. Hasman, M. L. Brongersma, "Dielectric gradient metasurface optical elements," Science, **345** (2014).

[26] Y. Chen, X. Li, Y. Sonnefraud, A. I. Fernandez-Domınguez, X. Luo, M. Hong, S. A. Maier, "Engineering the phase front of light with phase-change material based planar lenses," Sci. Rep., **5:8660**, pp. 1-7 (2015).

[27] A. V. Kildishev, A. Boltasseva, and V. M. Shalaev, "Planar photonics with metasurfaces", Science **339**, 1232009 (2013).

[28] A. Arbabi, Y. Horie, A. J. Ball, M. Bagheri, A. Faraon, "Subwavelength-thick lenses with high numerical apertures and large efficiency based on high-contrast transmitarrays," Nat. Commun., **8069** (2015).

[29] F. Aieta, M. A. Kats, P. Genevet, F. Capasso, "Multiwavelength achromatic metasurfaces by dispersive phase compensation," Science, **347**, pp. 1342-1345 (2015).

[30] M. Decker, I. Staude, M. Falkner, J. Dominguez, D.N. Neshev, I. Brener, T. Pertsch, Y.S. Kivshar, "High-Efficiency Dielectric Huygens' Surfaces," Adv. Opt. Mater., **3**, pp. 813–820 (2015).

[31] F. J. Garcia-Vidal, L. Martin-Moreno, T. W. Ebbesen, L. Kuipers, "Light passing through subwavelength apertures", Rev. Mod. Phys., **82**, pp. 729-787 (2010).

[32] F. Medina, F. Mesa, D. C. Skigin, "Extraordinary transmission through arrays of slits: a circuit theory model," IEEE Trans. Mic. Theory Tech., **58**, pp. 105-115 (2010).

[33] Z. Liu, E. Gao, F. Zhou, "Sharp multiple-phase resonances in a plasmonic compound grating with multislits," IEEE Photonics J., **10** (2018).

[34] D. Berry, R. Malech and W. Kennedy, "The reflectarray antenna," IEEE Trans. Antennas Propag., **11**, pp. 645-651 (1963).

[35] A. Díaz-Rubio, S. A. Tretyakov, "Acoustic metasurfaces for scattering-free anomalous reflection and refraction," Phys. Rev. B, **96** (2017).

[36] A. Epstein and G. V. Eleftheriades, "Passive lossless Huygens' metasurfaces for conversion of arbitrary source field to directive radiation," IEEE Trans. Antennas Propag., **62**, pp. 5680–5695 (2014).

[37] A. Epstein and G. V. Eleftheriades, "Floquet-Bloch analysis of refracting Huygens metasurfaces," Phys. Rev. B, **90**, p. 235127 (2014).

[38] J. P. S. Wong, A. Epstein, and G. V. Eleftheriades, "Reflectionless wide-angle refracting metasurfaces," IEEE Antennas Wireless Propag. Lett., **15**, pp 1293-1296 (2016).

[39] B. H. Fong, J. S. Colburn, J. J. Ottusch, J. L. Visher, D. F. Sievenpiper, "Scalar and tensor holographic artificial impedance surfaces," IEEE Trans. Antennas Propag., **58**, pp. 3212-3221 (2010).

[40] R. F. Harrington, D. T. Auckland, "Electromagnetic transmission through narrow slots in thick conducting screens," IEEE Trans. Antennas Propag., **28** (1980).

[41] L. Brekhovskikh, "Waves in Layered Media", Academic Press (1976).

[42] A. Epstein and G. V. Eleftheriades, "Arbitrary power-conserving field transformations with passive lossless omega-type bianisotropic metasurfaces," IEEE Trans. Antennas Propag., **64**, pp. 3880-3895 (2016).

[43] https://www.mathworks.com/

[44] V. S. Asadchy, M. Albooyeh, S.N. Tcvetkova, A. Díaz-Rubio, Y. Ra'di, and S. A. Tretyakov, "Perfect control of reflection and refraction using spatially dispersive metasurfaces," Phys. Rev. B, **94**, p. 075142 (2016).

[45] R. F. Harrington, "Time Harmonic Electromagnetic Fields", McGraw Hill, New York (1961).

[46] R. A. Depine, A. Lakhtakia, "Plane-wave diffraction at the periodically corrugated boundary of vacuum and a negative-phase-velocity material," Phys. Rev. E, **69** (2004).

[47] The terms "homogeneous" and "inhomogeneous" are algebraic concepts and unrelated to the "homogenization" of a discretized metasurface.

[48] https://www.cst.com/